\documentclass[]{aa}  
\usepackage{hyperref}
\usepackage[nolist]{acronym}
\acrodef{E6}[E6]{\emph{Experiment 6}}
\acrodef{ACR}[ACR]{anomalous cosmic ray}
\acrodef{GCR}[GCR]{galactic cosmic ray}
\acrodef{ESA}[ESA]{European Space Agency}
\acrodef{NASA}[NASA]{National Aeronautics and Space Administration}
\acrodef{IMP}[IMP]{\emph{Interplanetary Monitoring Platform}}
\acrodef{GME}[GME]{Goddard Medium Experiment}
\acrodef{SSD}[SSD]{silicon semiconductor detectors}
\acrodef{GEANT}[GEANT]{GEometry And Tracking}
\acrodef{SEP}[SEP]{solar energetic particle}
\acrodef{GLE}[GLE]{\emph{Ground Level Enhancement}}
\acrodef{KET}[KET]{\emph{Kiel Electron Telescope}}
\acrodef{HET}[HET]{\emph{High Energy Telescope}}
\acrodef{ACE}[ACE]{\emph{Advanced Composition Explorer}}
\acrodef{CRIS}[CRIS]{\emph{Cosmic Ray Isotope Spectrometer}}
\acrodef{PIN}[PIN]{Particle Identification Number}
\acrodef{EPHIN}[EPHIN]{\emph{Electron Proton Helium INstrument}}
\acrodef{SOHO}[SOHO]{\emph{SOlar and Heliospheric Observatory}}
\acrodef{HMF}[HMF]{heliospheric magnetic field}
\acrodef{LIS}[LIS]{local interstellar spectrum}
\acrodef{MeV}[MeV]{megaelectronvolt}

\begin{document}

   \title{Galactic cosmic ray hydrogen spectra and radial gradients in the inner heliosphere measured by the HELIOS Experiment 6}

\titlerunning{Experiment 6 GCR hydrogen spectra}

   \author{J. Marquardt
          \and
          B. Heber}
   \institute{Christian-Albrechts-Universität zu Kiel\\
              \email{marquardt@physik.uni-kiel.de}}

   \date{\today}

 
  \abstract
   {The HELIOS solar observation probes provide unique data regarding their orbit and operation time. One of the onboard instruments, the \ac{E6}, is capable of measuring ions from 4 to several hundred $\mathrm{MeV/nuc}$.}
   {In this paper we aim to demonstrate the relevance of the \ac{E6} data for the calculation of \ac{GCR}, \ac{ACR}, and \ac{SEP} fluxes for different distances from the sun and time periods}
   {Several corrections have been applied to the raw data: determination of the Quenching factor of the scintillator, correction of the temperature dependent electronics, degradation of the scintillator as well as the effects on the edge of semi-conductor detectors.}
   {Fluxes measured by the \ac{E6} are in accordance with the force field solution for the \ac{GCR} and match models of the anomalous cosmic ray propagation.  \ac{GCR} radial gradients in the inner heliosphere show a different behaviour than in the outer heliosphere}
   {}

   \keywords{\ac{GCR} --
                dE/dx-C-measurement principle--
                HELIOS
               }

   \maketitle
%
\section{Introduction}
\acresetall
In October 2011, the \ac{ESA} announced the selection of Solar Orbiter as one of the Cosmic Vision M missions, with the launch envisioned for 2019/2020. On August 12, 2018  the \ac{NASA} Parker Solar Probe was launched and reached its first perihelion on November 9, 2018. Thus, we have again spacecraft that determine in-situ the properties and dynamics of plasma, fields, and particles in the inner heliosphere. \citet{Ng-etal-2016} showed recently a solar cycle variation of 1-10~GeV $\gamma$-rays measured by the Fermi satellite, which is caused by \ac{GCR} particles interacting with the solar atmosphere. In order to investigate such temporal evolution it is worthwhile revisiting the energetic particle measurements by the HELIOS \ac{E6} performed in the 1970s within 0.4~AU in the light of advanced analysis and modelling techniques.

It has been recently shown that the \ac{E6} instrument is capable of measuring  the distribution of \ac{ACR} and \ac{GCR} ions from carbon ($z=6$) to silicon ($z=14$) in the energy range from a few \ac{MeV}/nucleon to several tens of \ac{MeV}/nucleon in the inner heliosphere during solar minimum \citep{Marquardt-etal-2018}, resulting in the first measurement of the radial gradient of anomalous oxygen within the Earth orbit. \citet{Bialk-1996} and \citet{Droege-1999} showed that the energy range of the instrument can be extended to above several 100 MeV/nucleon, allowing us to determine the energy spectra and the radial gradient of \acp{GCR}' hydrogen  in the inner heliosphere from 0.3 to 1 AU. 

    \begin{figure}
   \centering
   \includegraphics[width=0.8\columnwidth]{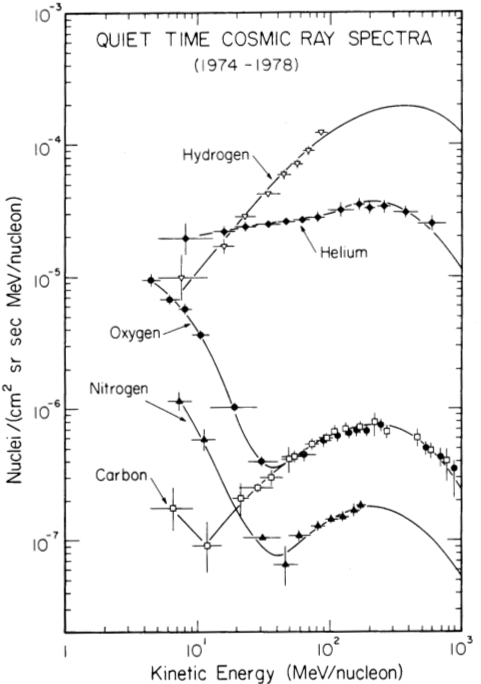}
   \caption{Quiet-time H, He, C, N, and 0 energy spectra measured at 1 AU over the period from 1974 to 1978. The \ac{ACR} component is reflected in the spectra by an enhancement at low energies for He, N, and 0 \citep[Fig. 1.2 of][]{Christian-1989}. The \ac{GCR} component dominates at energies above 30~MeV/nucleon for N and O and 50~MeV/nucleon for He, respectively} 
   \label{fig:acr}
   \end{figure}

\acp{GCR} encounter a turbulent solar wind with the embedded \ac{HMF} when entering the heliosphere. This leads to significant global and temporal variations in their intensity and in their energy as a function of position inside the heliosphere. This process is identified as the solar modulation of \acp{GCR} \citep[see for example][and references therein]{Potgieter-2013}. The analysis of the radial gradient of \ac{ACR} oxygen in the inner heliosphere within 0.5~AU by \citet{Marquardt-etal-2018} shows the need to improve particle transport models towards the Sun.

In what follows we show that the measurement capabilities of HELIOS \ac{E6} allow us to determine the hydrogen spectra up to above 800~MeV/nucleon. Figure~\ref{fig:acr} from \citet{Christian-1989} displays the quiet-time energy spectra for H, He, C, N, and O taken during quiet times from 1974 to 1978 1~AU by \ac{IMP}~8. We validate our results against the \ac{GCR} hydrogen measurements shown there. The accuracy of the instrument allows us to give upper limits of the radial gradient that are consistent with the ones reported by \citet{McDonald-etal-1977} and \citet{Webber-etal-1981} between 1 and 4.5~AU. 

\section{Instrumentation}
HELIOS A and HELIOS B were launched on December 10, 1974 and January 15, 1976, respectively. The two almost identical spinning space probes were sent into ecliptic orbits around the Sun. The orbital period around the Sun was 190 days for HELIOS A and 185 days for HELIOS B, and their perihelia were 0.3095 AU and 0.290 AU, respectively.
        \begin{figure}
   \centering
   \includegraphics[width=1\columnwidth]{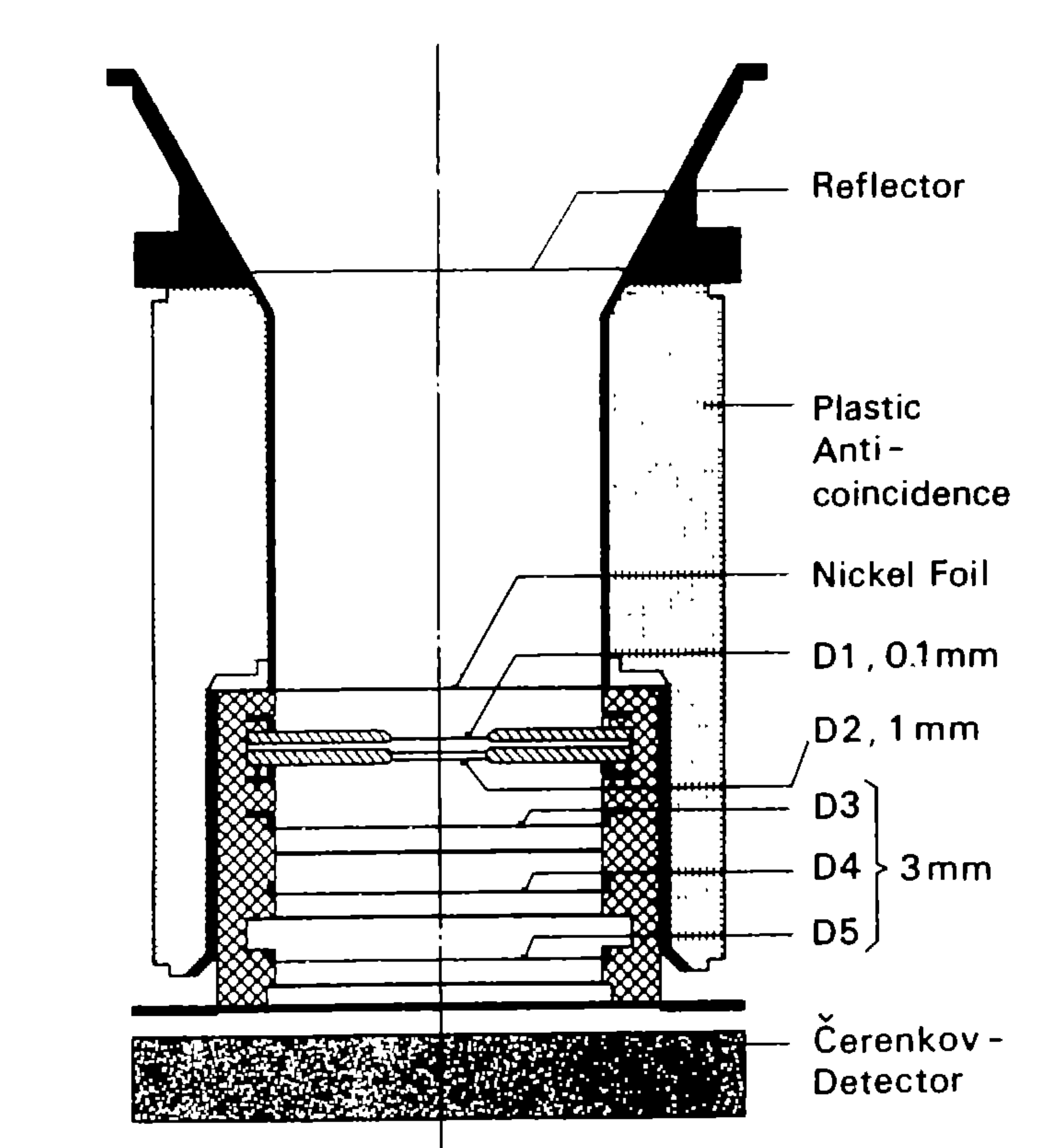}
   \caption{Schematic of the \ac{E6} detector setup.} 
   \label{E6}
   \end{figure}

A sketch of the \ac{E6} sensor is shown in Fig.~\ref{E6}. It consists of a stack of five \ac{SSD} (D1 to D5) and one Sapphire Cerenkov detector surrounded by a plastic anti-coincidence detector. The five \ac{SSD}s function as a ''standard'' $dE/dx-E$ telescope \citep{Brunstein-1964} with the Cerenkov detector used as anti-coincidence \citep[see for example][and references therein]{Marquardt-etal-2018} allowing us to measure hydrogen to silicon energy spectra in the energy range from a few to several tenths of MeV/nucleon. This method is based on at least two energy deposits, one in a thin detector transmitting ($dE/dx$) and another one in a thick detector stopping the incident particle ($E$) \citep[more details can be found in][and references therein]{Marquardt-etal-2014}. At energies above $\sim$50~MeV protons trigger the sapphire Cerenkov detector. In order to increase the geometric factor, both detectors D1 and D2 are not required for a valid coincidence. These integral channels are called P51 for protons and A48 for heavier ions and the identification of ions is based on the $dE/dx-dE/dx$  and $dE/dx-C$ method \citep{Kuehl-etal-2016, Linsley-1955}. 

\begin{figure}
   \centering
   \includegraphics[width=1.1\columnwidth]{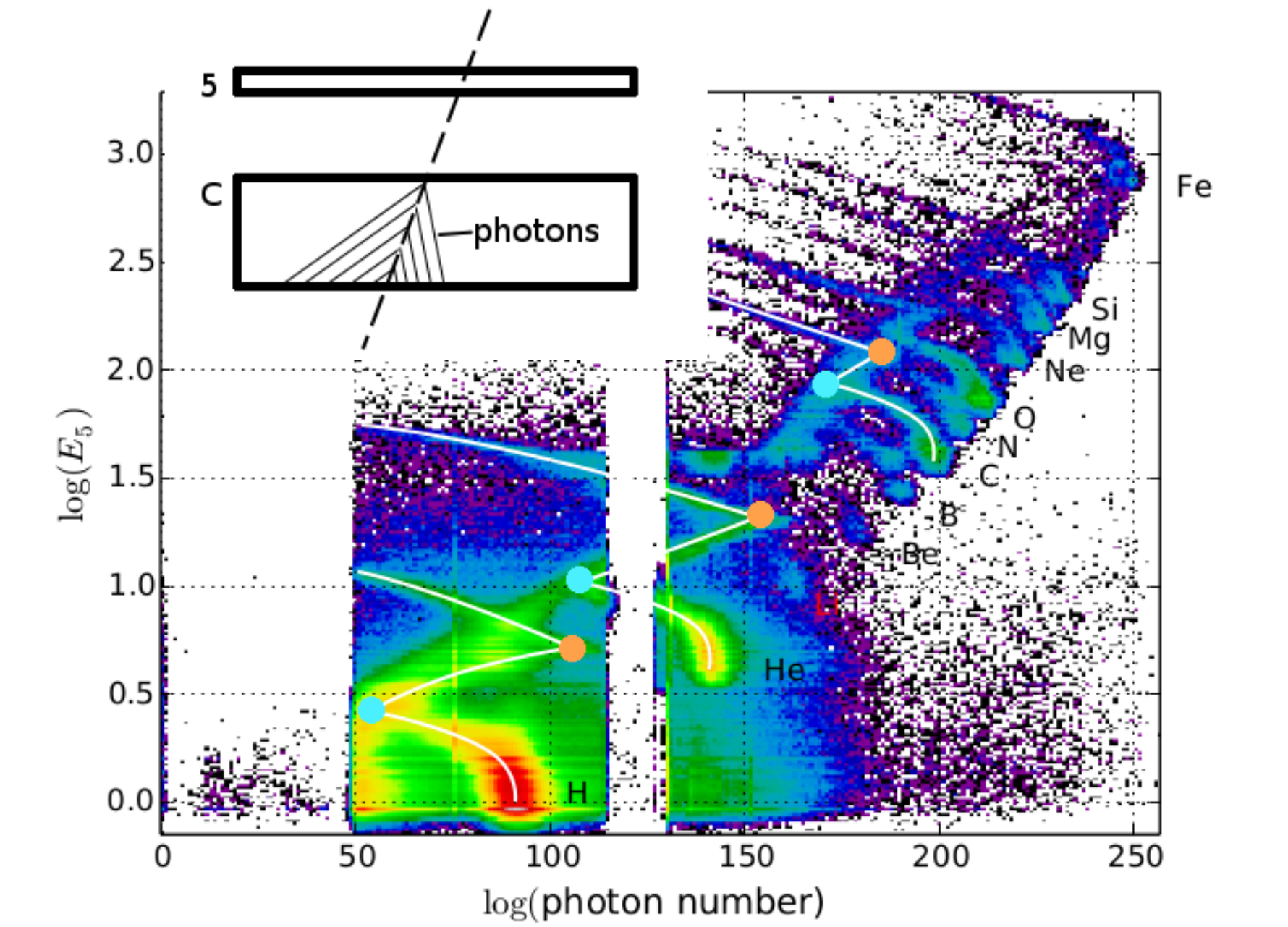}
   \caption{Integral channel of the HELIOS \ac{E6}. The orange point marks the penetration of the Cerenkov (C) detector, the blue point marks the exceeding of light speed inside the medium resulting in the particles emitting Cerenkov-Radiation}
              \label{p51_all}%
    \end{figure}

The $dE/dx-dE/dx$ method is based on the energy loss in two detectors allowing us to identify different particle species in certain energy ranges.  However, this method has two major disadvantages, which are (1) some areas of the two dimensional energy loss plane are populated by different elements and (2) the signal from particles that penetrate the instrument from the back cannot no longer be distinguished from the ones that penetrate the instrument from the front. By adding a Cerenkov detector the overlap of different species can be minimized and one can discriminate against backward penetrating particles. This so called $dE/dx-C$-method \citep{Linsley-1955} is applied to charged particles that completely penetrate a semi-conductor detector 5 and a Cerenkov detector C, which is placed underneath (see inset in Fig.~\ref{p51_all}). If they penetrate C faster in the dielectric material than light can propagate, they produce a measurable light flash (Cerenkov radiation). The threshold speed of $v > c/n$ depends on the refractive index  $n$ of the material. Plotting  the energy-loss by ionization, $\Delta E$ in A, as a function of the Cerenkov detector signal results in characteristic curves, clearly separated for different atomic numbers, with their slopes depending on particle speed. Thus, the method allows an identification of the penetrating particles and a determination of their energy above a threshold speed. 
Figure~\ref{p51_all} shows measurements  by Helios \ac{E6}, where the Cerenkov detector is made of sapphire, which is also a scintillator responding to the ionization energy loss of the particle in the detector. The different ion  tracks are identified in the figure and the orange and blue circles mark those points along the track where the particles penetrate the sapphire and where the Cerenkov light production starts, respectively. 
Thus, the $dE/dx-dE/dx$ method is used along the tracks starting at the orange point and ending at  the blue point, and the $dE/dx-C$ method after the blue point.  As is evident from Fig.~\ref{p51_all}, charged particle measurements can suffer from various imperfections. Therefore, modelling of the physical processes and of the instrument geometry, as well as the environment, is essential to understand such measurements \citep[e.g.][]{Heber-etal-2005, Kuehl-etal-2015,Marquardt-etal-2014}. 

\section{\acl{E6} modelling}
In order to understand the Helios \ac{E6} response to penetrating ions, a \ac{GEANT}~4 simulation \citep[][]{Agostinelli-etal-2003} has been setup that has to include optical photon tracking as well as Birk's quenching \citep{Birks1951} in the sapphire detector, as discussed in what follows.

While usual anorganic scintillation counters reach a typical scintillation yield of one photon per 100 eV deposited energy, the sapphire Cerenkov detector scintillates with an efficiency of one photon per 50 keV deposited energy. The reason for this is the self-absorption of the emitted light inside the scintillator and the emitted photons being of higher energy than the photon energy at which the photo-multiplier reaches peak efficiency. In common anorganic scintillators those effects are bypassed by doping the base material.
Due to the low scintillation efficiency of the detector, the light output from scintillation falls in the same order of magnitude as the light output from Cerenkov radiation. Otherwise it wouldn’t be possible to measure the Cerenkov effect and scintillation light with the same detector. The sum of the emitted photons can be seen in Fig. \ref{p51_all}. Cerenkov radiation is emitted as soon as particles have a higher speed than light in the medium; in the case of the sapphire  $v_n=V_0/n=0.566\cdot$c with $n=1.77$  has been used. Cerenkov radiation is always emitted anisotropically while scintillated photons are isotropic.

\begin{figure}
   \centering
   \includegraphics[width=1.0\columnwidth]{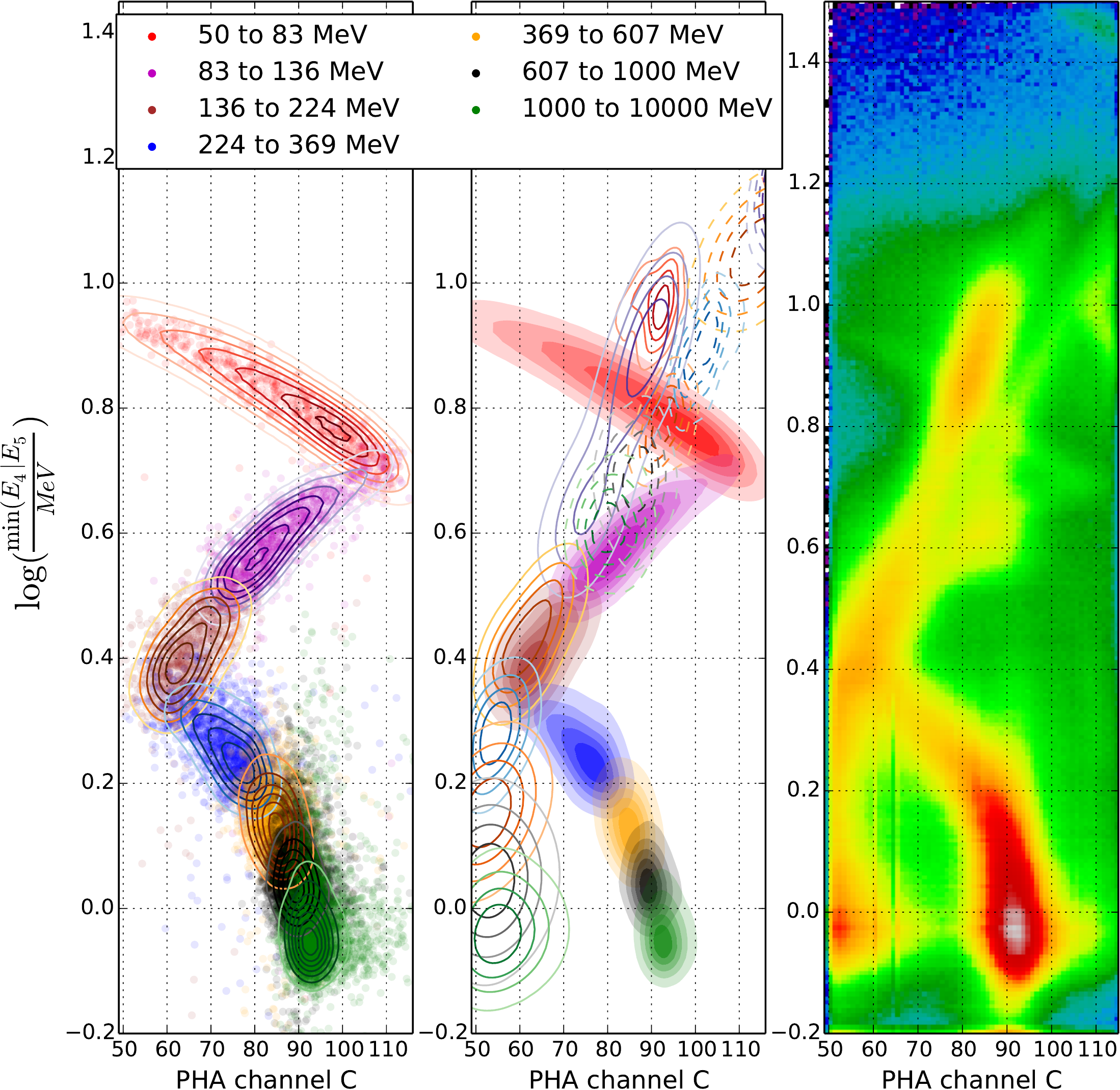}
   \caption{Simulated protons in the integral channel. We colour-coded the different input energy ranges for the simulation; they are logarithmic equidistant except for the last integral bin. The lines on the middle panel are for backwards protons and the dashed lines for backwards He. On the right panel, real measurements are shown for comparison. All energies are in MeV/nuc}
              \label{p51_response}%
    \end{figure}

\begin{figure}
   \centering
   \includegraphics[width=1.0\columnwidth]{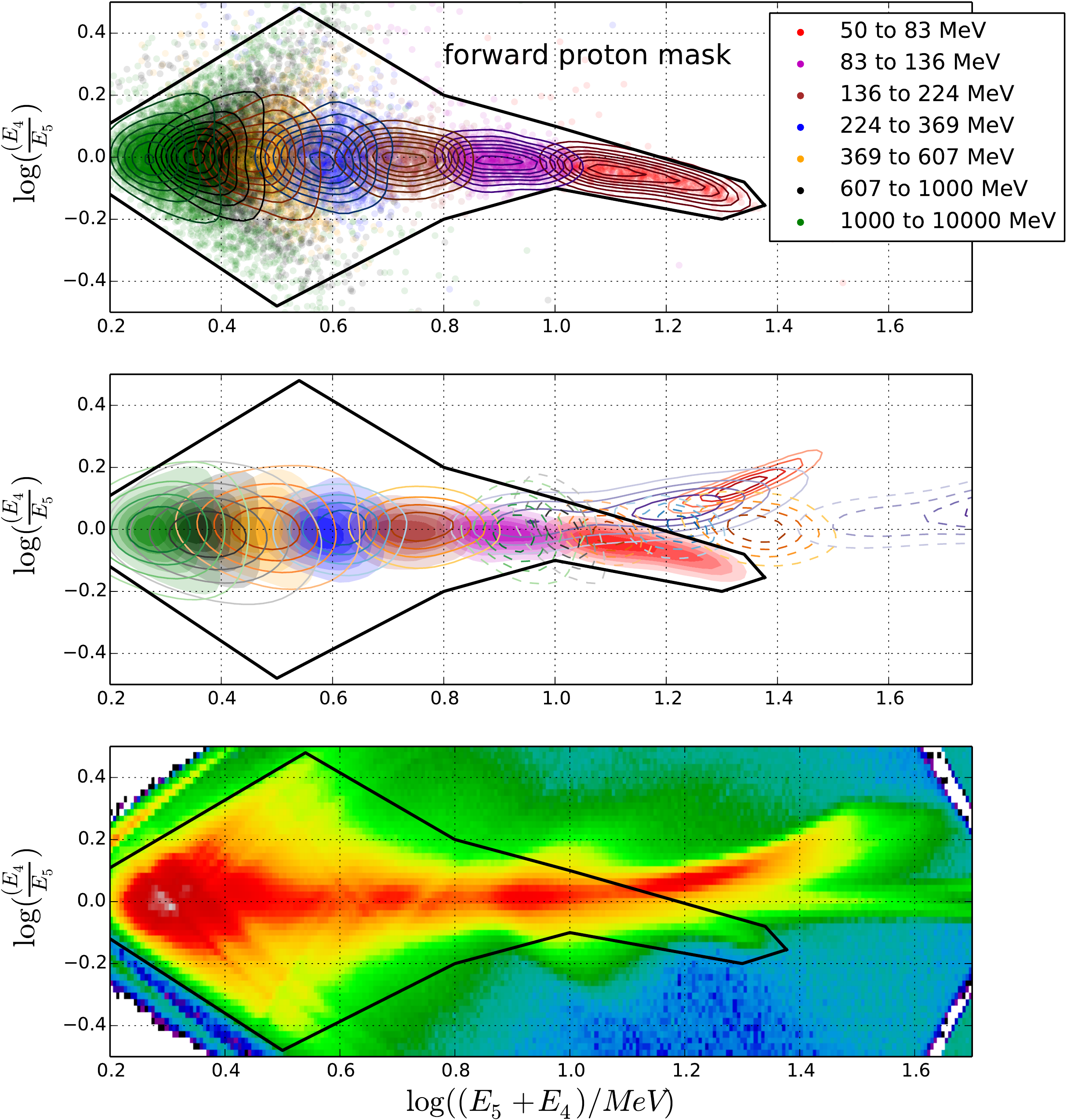}
   \caption{Simulated protons in the integral channel. We colour-coded the different input energy ranges for the simulation; they are logarithmic equidistant except for the last integral bin. The lines on the middle panel are for backwards protons and the dashed lines for backwards He. On the lowest panel real measurements are shown for comparison. All energies are in MeV/nuc}
              \label{p51_response_1}%
    \end{figure}

In Fig. \ref{p51_all} it is also noticeable that neither the orange nor the blue points align. This is due to quenching \citep{Birks1951}. The higher the energy deposit per path length, $\frac{dE}{dx}$, the lower the number of photons per energy deposit. Furthermore, the upper side of the detector C has been blackened to avoid the reflection of light. For speeds much larger than $v_n$ the light output is dominated by Cerenkov light that reflects the direction of the incoming particles. Thus, Cerenkov light from particles entering the detector from behind gets absorbed, while photons from the scintillation process are still counted. This leads to a separation of forward and backward penetrating particle tracks in Fig.~\ref{p51_all}. However, particles with speed $v<v_n$ lead to a photon distribution that is isotropic resulting in insufficient discrimination between forward and backward particles below 0.566$\cdot$c. In order to improve the rejection of backwards penetrating ions, we calculated the expected distributions for forward and backward penetrating protons as well as the ones for backward penetrating helium in Figs. \ref{p51_response} and \ref{p51_response_1}. In all our simulations, quenching in the sapphire detector has been taken into account, by using Birk's formula
\begin{equation}
   \frac{dL}{dx}=S\frac{\frac{dE}{dx}}{1+k_B\frac{dE}{dx}}
,\end{equation}
using as parameter $S = \frac{20}{MeV}$ and $k_B = 50\cdot 10^{-6}\frac{mm}{MeV}$.

   \begin{figure}
   \centering
   \includegraphics[width=1.1\columnwidth]{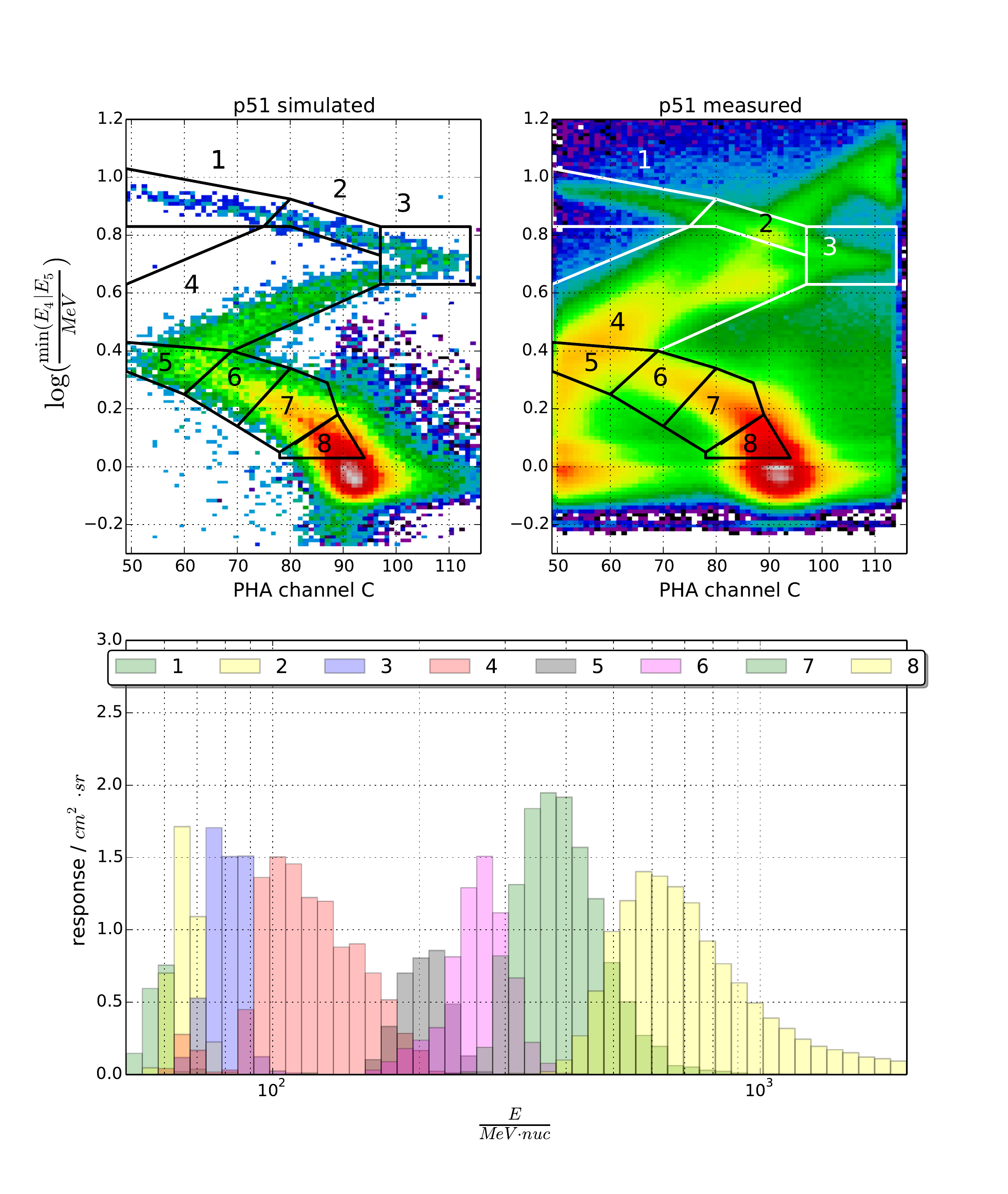}
   \caption{Simulated protons on the left and corrected measured protons on the right in the integral channel. Bottom: Responses for the boxes shown in upper panels}
              \label{p51_response_2}%
    \end{figure}

\begin{figure}
    \centering
   \includegraphics[width=1.1\columnwidth]{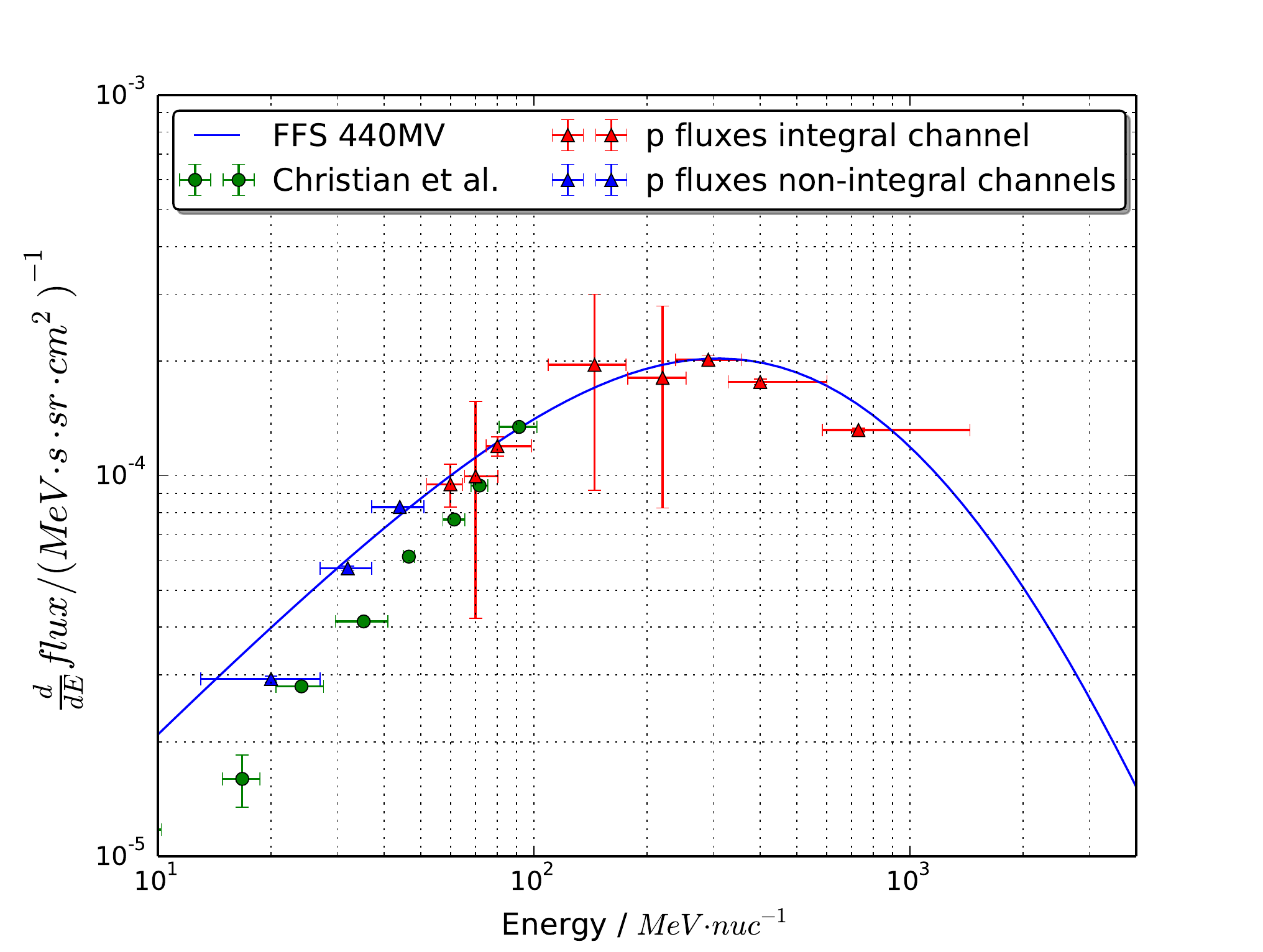}
   \caption{Blue and red symbols show the Helios proton fluxes from the energy channels for stopping and penetrating protons, respectively. The green symbols are the fluxes measured by \ac{IMP}~8 taken from Fig.~\ref{fig:acr}. The blue line shows the force field solution utilizing a modulation parameter $\Phi=440$~MV. Details can be found in the text.} 
   \label{p51_fluxes}
   \end{figure}

Taking the above-mentioned effects into account, we performed a simulation with one~Billion protons in the energy range from 40~MeV/nucleon up to 10~GeV/nucleon impinging isotropically on the \ac{E6} sensor. Results for protons that cross the sensor from the front are summarized in the left panel of Fig.~\ref{p51_response} and in the top panel of Fig.~\ref{p51_response_1}. The first of the two figures displays the minimum logarithmic energy loss $\Delta E$ in \ac{SSD}~4 and \ac{SSD}~5 as a function of the light output of the Cerenkov detector C in six different energy bands from 50 to 83~MeV/nucleon (red contour lines) to 607 to 1000~MeV/nucleon (black countour lines). The width of these channels is chosen so that they are spaced equally in the logarithm of the energy boundaries. An integral channel from 1000 to 10000~MeV is shown by the green contour lines.  We note that $min(\Delta E_4, \Delta E_5)$ results in a sharper pattern since it minimizes the stochastic nature of the energy deposition \citep[see also][]{Kuehl-etal-2015}. The right panel of Fig.~\ref{p51_response} displays the corresponding quiet-time measurements obtained from December 1974 to July 1977. By comparing both panels with each other, we find that the calculated track reflects significant features in the measurements as there are the position of the turning points when crossing the detector and the onset of the Cerenkov effect. However, we find some significant features that must be caused by backward penetrating helium and protons. In order to make use of all the information available to us, the upper panel of Fig.~\ref{p51_response_1} displays the position of the energy intervals in the difference of the logarithmic energy losses in \ac{SSD}~5 and \ac{SSD}~4 versus the sum of the logarithmic energy losses in \ac{SSD}~4 and \ac{SSD}~5 matrix. Although the energy losses converge above a certain energy (here above $\sim$~83~MeV), differences are found for lower energies. The computed distribution can therefore be used to define a mask with all valid entries for forward penetrating protons. The middle panels of both figures display in addition the distribution for protons and helium penetrating the instrument from the back indicated by the solid and dashed contour lines, respectively. The different colours of the contour lines give the incoming energy range of the backward penetrating particles in MeV/nucleon. Comparing the middle panel and the right panel of Fig.~\ref{p51_response} and the middle and the lower panel of Fig.~\ref{p51_response_1}, all simulated features are seen in the in-flight matrix indicating that the simulation reflects the measurements very well. From both figures one notes that the contour lines for backward penetrating protons below 83 MeV are well outside the mask (Fig.~\ref{p51_response_1}) and the forward penetrating proton track in Fig.~\ref{p51_response}. A significant reduction can even be obtained up to 136~MeV protons (cyan contour lines). In the energy range from 136 to 230~MeV backward penetrating and forward penetrating protons cannot be distinguished. Above 230~MeV the Cerenkov effect sets in and the tracks in Fig.~\ref{p51_response} separate again. The dashed contour lines in the middle panel of both figures show the distributions for backward penetrating helium. The mask defined in Fig.~\ref{p51_response_1} rejects backward penetrating helium with energies lower than $\sim$350~MeV/nucleon (blue dashed contour lines). At energies above this threshold forward penetrating protons and helium cannot be distinguished. In Fig.~\ref{p51_response_2} (right panel) we applied our mask to the in-flight measurements in $min(\frac{dE}{dx}_{SSD5},\frac{dE}{dx}_{SSD4})-C$ distribution.  Although we retrieve a significant reduction of the contribution of backward penetrating particles, we are left with areas that cannot be cleaned. To obtain energy spectra we defined eight boxes as shown in the upper left and right panel in Fig.~\ref{p51_response_2} for the simulated and measured matrix, respectively. The boxes were defined as a compromise between equal logarithmic energy spacing and splitting the different particle populations. Boxes 1 and 3 were chosen to completely avoid contamination from backward penetrating particles. Box 2 only contains backward penetrating helium. Box 4 was chosen to contain the parts of the spectra in which separation of forwards and backwards protons as well as backwards helium are impossible to distinguish from one another. Box 5 is free of helium and the energy of forwards and backwards protons is roughly the same, allowing for easier statistical separation. Boxes 6 and 7 are again free of contamination and spaced equally. Box 8 has a sharp cut-off to avoid electrons and near relativistic protons from entering the box. We note that electrons play a very minor role as they will be always minimally ionizing and at near-light speed and are thus below and to the right of box 8. The lower panel displays the computed forward penetrating proton response functions $R^{i=p}_{\alpha}(E)$ as a function of the kinetic energy $E$ for each box ranging from about 50~MeV to above 2~GeV. Following \citet{Sullivan-1971} the measured count rate $C_i$ for each channel is given by
\begin{equation}
    C_i = \sum_{\alpha = 1}^{n} \int_{0}^\infty R{^i}_\alpha (E) J_\alpha (E) dE
\label{eq:S1} ,\end{equation}
with $\sum_\alpha$ being the sum over all particles species contributing to each channel $C_i$, and $J_\alpha (E)$ being the energy spectrum for each particle species. The total contribution is then given by the integral over all possible energies. For an ideal detector, that is a detector that is only sensitive to one particle type with a response function that is constant $R_i$ in the energy range from $E_l$ to $E_u$ and otherwise zero:
\begin{equation}
     R(E)= \left\{ 
    \begin{array}{ll}
0 & \textrm{for } 0<E<E_{l}\\
R_i & \textrm{for } E_{l}\leq E \leq E_{U}\\
0 & \textrm{for } E>E_{u}\\
\end{array}\right.
\label{eq:S2}
.\end{equation}
In that case, Eq.~\ref{eq:S1} reduces to
\begin{equation}
    C_i = R_i \cdot (E_u - E_l) \cdot I (\left< E \right> ),
\end{equation}
where $\left< E \right>$ is the mean energy of channel $i$ and $J_i(\left< E \right> )$ can be easily computed by $$I (\left< E \right> )= \frac{C_i}{R_i.\cdot (E_u - E_l)}\; .$$ 
Although the response function for each box is deviating from the ideal ones described by Eq.~\ref{eq:S2}, we  approximate $R \cdot (E_u - E_l)$ by the integral of the response function $\int\limits_{E_l}^{E_u} R_i(E) dE$ and $\left< E\right>$ is the energy $E$ for which the response function has a maximum.  The results are visualized in Fig. \ref{p51_fluxes}. In this figure we added three channels for stopping protons to extend the energy range down to about 10~MeV \citep[see][and references therein]{Marquardt-etal-2018}. While the y errors account for statistical errors only, the x errors mark the energies when the response has been decreased to $\frac{1}{6}$ of the maximum response of each box. This simple method has been chosen since it shows in an intuitive way the results applicable to a response function that has a box or a gaussian shape, respectively. For comparison the green symbols display the hydrogen measurements from Fig. \ref{fig:acr}. Taking into account the different measurement times from 1974 to 1978 for \ac{IMP} 8 and from the end of 1974 to 1977 for Helios A, the agreement between both data sets is remarkably good. Taking these uncertainties into account, our analysis shows that the \ac{E6} can be utilized to determine proton energy spectra in the range from 10 to 50~MeV from energy channels of stopping particles and from 60 to about 600~MeV for penetrating particles.

\begin{figure}[h]
   \centering
   \includegraphics[width=1.1\columnwidth]{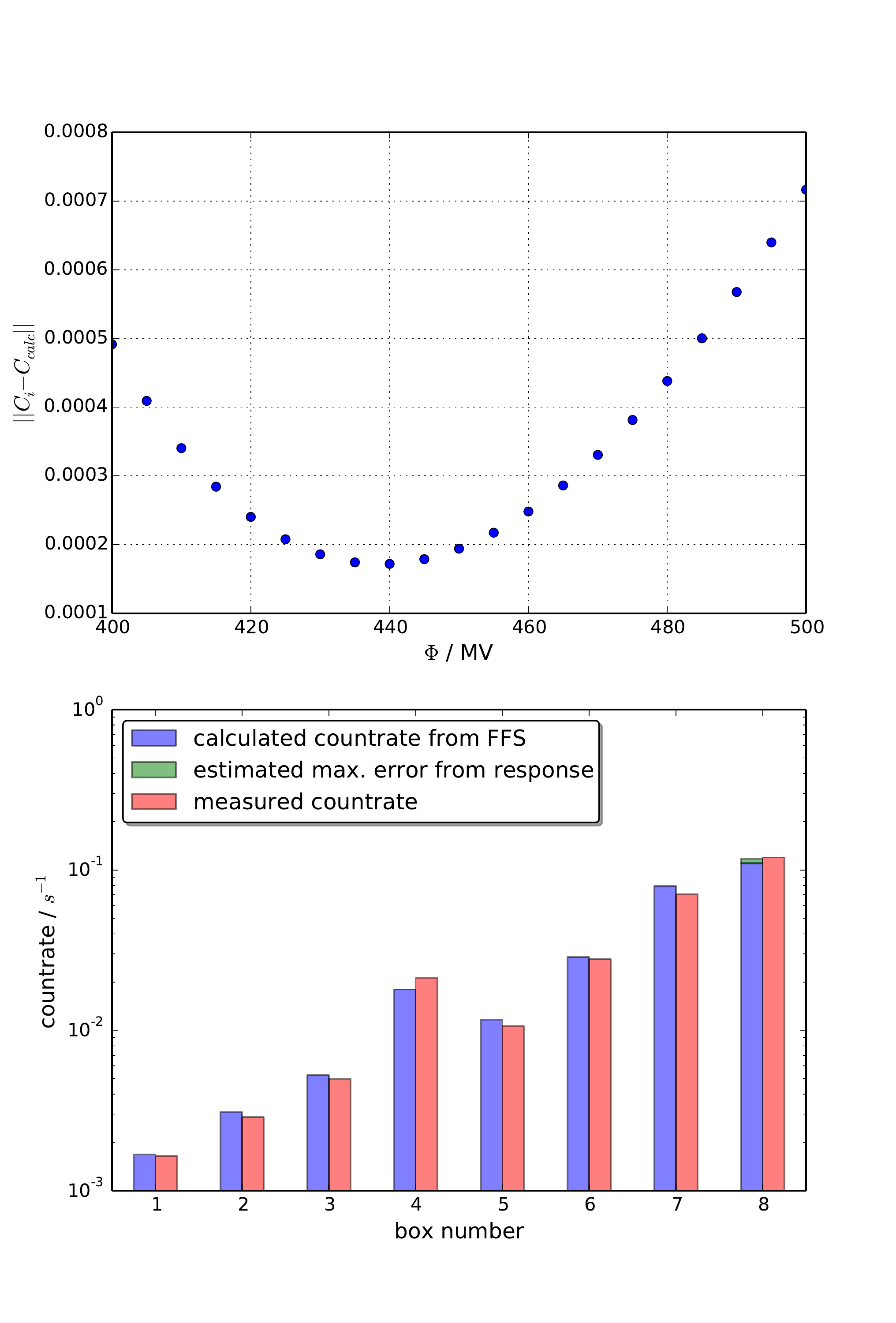}
   \caption{Upper panel: Norm of the difference between calculated and measured count rates as a function of the modulation parameter $\phi.$ Lower panel: Proton count rates measured and calculated from FFS seen in Fig. \ref{p51_fluxes}.} 
   \label{p_c}
\end{figure}

During quiet times the energy spectra of protons can be approximated by the force field solution \citep[FFS, see][and references therein]{Gleeson-Axford-1968, Caballero-Lopez-Moraal-2004}. As \ac{LIS} we used the one given by \citet{Burger-etal-2000},
\begin{equation}
     J_i(\Phi) = J_{LIS,i}(T+\Phi)\frac{T+2E_{0,i}}{(T+\Phi)(T+\Phi+2E_{0,i})}
,\end{equation}
where $\Phi$=$(Ze/A)\phi $ is the modulation function and  $\phi$ is the modulation parameter. Rewriting Eq. \ref{eq:S1} we can minimize the norm
\begin{equation}
     \|\sum\limits_{i=1}^{8} (C_i - \sum\limits_{\alpha = 1}^{n} \int_{0}^\infty R{^i}_\alpha (E) J_{1AU}^\phi (E) dE)\|
\end{equation}
in order to obtain the modulation parameter that fits the Helios \ac{E6} measurements best. Figure~\ref{p_c} (left panel) shows the norm as a function of the modulation parameter $\phi$ showing a minimum at $\phi=440$~MV. Using data from the neutron monitor network \citet{Usoskin-etal-2005} and \citet{Gieseler-Heber-2017} computed Bartels rotation averaged modulation parameters from 1951 to 2004 using the \ac{LIS} from \citet{Burger-etal-2000}. From 1974 to the end of 1978 and from the launch of Helios A in December 1974 and the end of 1977, $\phi$  varies from $\phi_{min}=404$ to $\phi_{max}=670$ and from $\phi_{min}=404$ to $\phi_{max}=494$, respectively. The mean values are  $\langle \phi \rangle = 474 \pm 28$ and $\langle \phi \rangle = 435 \pm 23$, respectively. The latter value compares well with the one found in our analysis. The right panel visualizes the distribution of the total measured count rates in comparison to the calculated ones. Since the response of  box 8 never reached 0 (see Fig.~\ref{p51_response_2}) we estimate the contribution from protons above 10 GeV by assuming that the response function is constant between 10 and 100 GeV. This extra contribution is shown by the green box on top of the blue one in Fig. \ref{p_c}. Thus we conclude that protons above 10 GeV play a very minor role. 

\section{Radial gradients}
\begin{figure}
     \centering
    \includegraphics[width=\columnwidth]{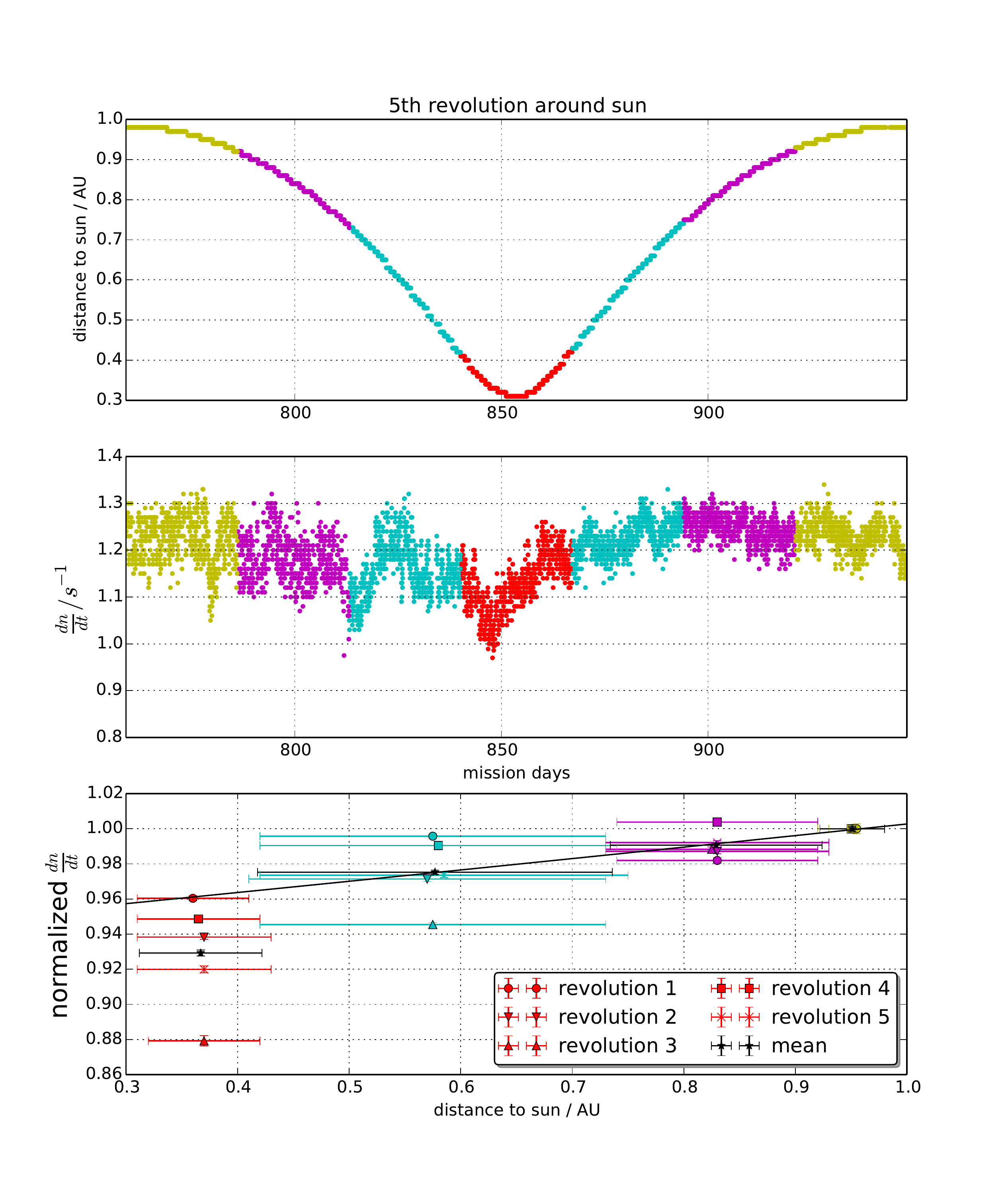}
    \caption{Top panel shows radial distance to sun versus time. We colour-coded the length of one Bartels rotation centred around the closest point to sun. The middle panel shows corresponding count rates. The lowest panel shows the averaged count rates for different revolutions versus the distance.}
    \label{fig:cr_grad}
\end{figure}

Another important open question is how cosmic rays are transported towards the Sun in the inner heliosphere. \citet{Marquardt-etal-2018} showed that anomalous cosmic ray oxygen penetrates deeper into the inner heliosphere as predicted by computations. \citet{Strauss-Potgieter-2010} and recently \citet{Ng-etal-2016} found solar cycle variation in the 1-10~GeV $\gamma$-rays measured by the Fermi satellite in the vicinity of the Sun. Thus in contrast to our current understanding, cosmic rays penetrate deeply into the Sun's corona. In order to advance our understanding, it is important to know the radial variation of the \ac{GCR} flux within 1~AU.  With the improved data analysis of the \ac{E6} experiment, we investigate in what follows the radial gradient of galactic cosmic ray protons in the energy range from about 250 to about 700~MeV, combining boxes 6 to 8. We investigate the radial variation using a two-step approach. Since the flux obtained in this channel results from the product of the integral channel and the number of entries in  boxes 6 to 8, we first determine the radial variation in the integral channel and then the one in the box channel. The integral channel is the channel that measures forward and backward penetrating protons and electrons and backward penetrating helium above 50 MeV/nucleon for ions and above 10~MeV for electrons.  Figure~\ref{fig:cr_grad} displays in the top and middle panels the radial distance to the Sun and the count rate in the integral proton channel for the fifth orbit of Helios 1 from December 30, 1976 (mission day 750) to July 18, 1977  (mission Day 950). Marked by different colours are Bartels rotation averages centred around the closest approach, allowing us to determine the radial dependence of this count rate. In the lowest panel, the mean of the Bartels average of the count rates prior and after closest approach are displayed as a function of radial distances for all orbits, which occurred during quiet times from launch of the satellite to July 18, 1977. In order to compare the different orbits to each other all count rates are normalized to the ones observed between 0.9 and 1~AU. Although we find a wide spread, a clear trend of decreasing flux with radial distance is obtained. In order to minimize the influence of temporal variations we average the normalized values for all five orbits. They are shown in Fig.~\ref{fig:cr_grad} by the black bullets. By fitting a line to the logarithms of the three outer bins we obtain a radial gradient of $6.6\pm 4$\%/AU. This value is consistent with the one obtained by \citet{Bialk-1996} but larger than the ones published by \citet{McDonald-etal-1977} and \citet{Webber-Lockwood-1981} summarized in Table~\ref{tab:gradient}. We note that the flux at 0.35~AU is much lower than the expected one from our fit. This is in agreement with the observation of the radial gradient of anomalous oxygen increasing in the inner heliosphere \citep{Marquardt-etal-2018}. 

\begin{figure}
    \centering
    \includegraphics[width=\columnwidth]{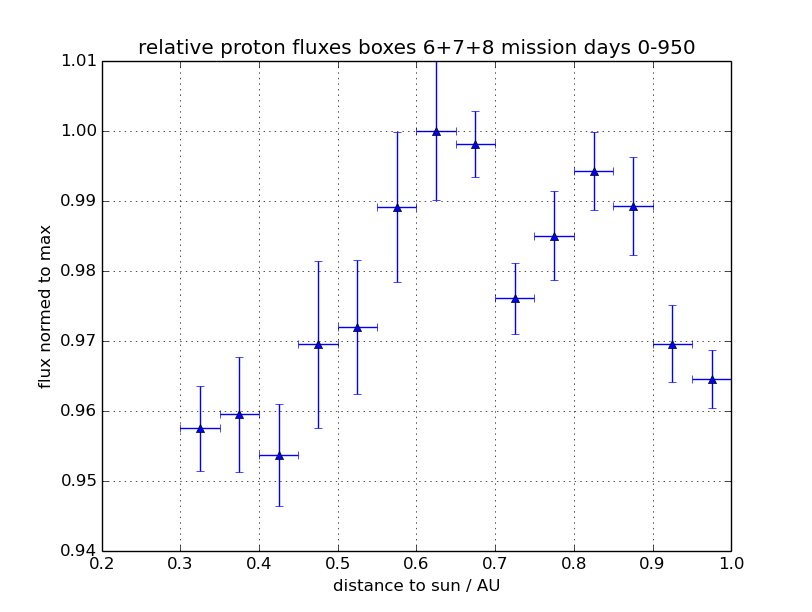}
    \caption{Flux in the energy range from 250 to 700~MeV as a function of radial distance using a bin width of 0.05~AU.  The values have been normalized to the maximum value at a distance of 0.6~AU. Details can be found in the text. }
    \label{fig:gradient}
\end{figure}

\begin{table}[]
     \caption{Selected radial gradients obtained in the heliosphere by \citet{McDonald-etal-1977}, \citet{Webber-etal-1981}, \citet{Bialk-1996}, and this study.} 
    \centering
\renewcommand{\arraystretch}{1.4}
    \begin{tabular}{|c|c|c|}
\hline
    Distance range     &  $G_r$ & Energy\\ 
    AU & \%/AU & MeV \\ \hline
    \multicolumn{3}{|c|}{From \citet{McDonald-etal-1977}}\\\hline
    1.25 - 4.2 & $4.1\pm 3.7$ & 210-275 \\
    1.25 - 4.2 & $2 \pm 4 $ & 275 -380 \\
    1.25 - 4.2 & $1.3 \pm 5 $ & 380 - 460 \\\hline
    1 - 3.8 & $0\pm 4$ & 210-275 \\
    1 - 3.8 & $2.5\pm 4$ & 275-380 \\
    1 - 3.8 & $3.8\pm 5$ & 380 - 460 \\ \hline
    \multicolumn{3}{|c|}{From \citet{Webber-Lockwood-1981}}\\\hline
    2 - 28 & $2.5\pm 0.5$ & $>60$ \\ \hline
    \multicolumn{3}{|c|}{This study}\\\hline
    0.4 - 1 & $6.6\pm 4$ & $>50$ \\\hline
    0.3 - 1 & $2\pm 2.5 $ & 250 - 700 \\\hline
    \end{tabular}
    \label{tab:gradient}
\end{table}
In the second step we used the same approach as for the integral channel for the differential proton channel sensitive to protons between 250 and $\sim 700$~MeV. Here we binned the data so that we get a radial resolution of 0.05~AU per bin as displayed in Fig.~\ref{fig:gradient}.
 The values have been normalized to the maximum value at a distance of about 0.6~AU.  We note that transient and recurrent Forbush Decreases \citep{Richardson-2004, Richardson+Cane-2011}, which are short term flux decreases in the cosmic ray flux, lead to larger variation (error) than the statistical ones. However, Fig.~\ref{fig:gradient} shows no clear overall trend. In order to estimate the radial gradient we need to minimize the influence of temporal effects. Therefore we divided the data set into measurements close to the Sun (0.3 to 0.6~AU) and far away from the Sun (0.7-1~AU), respectively. Our analysis leads to a radial gradient of $G_R=2\pm 2.5$\%/AU that is in good agreement with the one published by \citet{McDonald-etal-1977}, \citet{Bialk-1996}, and \citet{Webber-etal-1981}. Because of the limited \ac{E6} capabilities the uncertainties in the differential flux measurements do not indicate any increase of the radial gradient towards the Sun. Although the count rate profile of the integral channel as well as the anomalous oxygen indicate an increase of the radial gradient within 0.5~AU, only the measurements from the Parker Solar Probe will validate or disprove the Helios observations presented here. 
\section{Summary and conclusions}
The \acf{E6} aboard the Helios space probes was designed to measure ions and electrons in the energy range from a few MeV/nucleon to above 50~MeV/nucleon and 0.15 and above 10~MeV for electrons. In order to compute the proton energy spectrum above 50~MeV/nucleon, the instrument utilizes the $\frac{dE}{dx}-C$ method. A sophisticated model of the instrument has been developed on the basis of the \acf{GEANT}-4 package. We computed the response of the instrument not only to forward penetrating protons but also to hydrogen and helium that penetrate the sensor from behind. In order to reduce the background to these unwanted contributions, the energy loss distributions in the two silicon detectors have been evaluated. By adding a simple mask the background of backward protons below 130~MeV could be reduced significantly. For energies between 130 and 250~MeV, backward and forward penetrating protons cannot be distinguished from the signal of the last three detectors. At higher energies from above 250~MeV the Cerenkov effect sets in and forward and backward penetrating particle tracks separate again (see Fig.~\ref{p51_response_1}). Applying the ''background'' rejection derived from simulations an energy response (lower panel in Fig.~\ref{p51_response_2}) for different masks shown in the upper two panels of Fig.~\ref{p51_response_2} were computed. These response functions were used to compute the \ac{GCR} spectrum during quiet times from December 1974 to July 1977. The flux in each mask (box) was determined by applying a simple inversion. Taking into account the different measurement periods used in the study by \citet{Christian-1989} and our analysis, the spectra derived from Helios and \ac{IMP}~8 measurements agree very well with each other. Our analysis resulted in $\phi=440$~MV, which is in very good agreement with mean $\phi=435$~MV derived from the values published by \citet{Usoskin-etal-2005}. Thus we conclude that Helios \ac{E6} can be used to determine the proton spectra up to above 600~MeV. However, not only the intensity close to Earth can be determined but also the radial gradient within 1 AU. In contrast to \citet{Webber-Lockwood-1981} who determined a radial gradient of $2.5\pm 0.5$\%/AU between 2 and 28~AU, we found a radial gradient of $6.6\pm 4$\%/AU between 0.3 and 1~AU for above 50~MeV protons. Our analysis indicates an increasing radial gradient within 0.5~AU. The analysis from \citet{Bialk-1996} using an integral channel with energies above 135~MeV results in somewhat lower gradients. This trend is continued when we determine the radial gradient for protons in the energy range between 250 and 600~MeV protons to $2\pm2.5$\%/AU, which is in good agreement with the values found by \citet{McDonald-etal-1977} obtained between about 1 to about 4~AU. The Parker Solar Probe has explored the inner heliosphere on its first orbit during the same magnetic polarity of the Sun as in the 1970s and during solar minimum conditions. Therefore the results from the Parker Solar Probe will enable us to find out the following information: 
1) whether the radial gradient during the current solar cycle is consistent with the one obtained in the 1970s between 0.5 and 1 AU; 2) whether the radial gradient increases with decreasing distance to the Sun within 1 AU; and 3) in the event that the Parker Solar Probe results confirm the HELIOS results, we can ascertain the implications for cosmic ray propagation models.  

\bibliographystyle{aa}


\bibliography{papers}


\end{document}